\documentclass[letterpaper,11pt]{article}

\usepackage[T1]{fontenc}
\usepackage[utf8]{inputenc}
\usepackage{geometry}
\usepackage{graphicx}
\usepackage{amsmath}
\usepackage{pdfpages}
\usepackage[hidelinks]{hyperref}
\usepackage[
articletitle=true,
doi=true,
maxauthors=8
]{achemso}

\title{Towards a universal meta-optics solver via large language models}
\author{Huanshu Zhang,\textsuperscript{a} Lei Kang,\textsuperscript{a} Yuyan Chen,\textsuperscript{b} Luxiang Wang,\textsuperscript{b} Zhaolong Cao,\textsuperscript{b,+}\\ Douglas H. Werner\textsuperscript{a,*}}
\date{\textsuperscript{a} The Pennsylvania State University, Department of Electrical Engineering, University Park, PA 16802, USA\\
\textsuperscript{b} School of Electronics and Information Technology, Sun Yat-Sen University, Guangzhou 510275, China}

\begin{document}

\maketitle

\begin{abstract}
\noindent\textbf{Abstract.} Metasurface design increasingly requires fast models that can operate across structurally distinct device families, rather than retraining a separate surrogate for every geometry class. Conventional neural network surrogates often depend on fixed-dimensional descriptors, family-specific output formats, and repeated architecture tuning, which limits their scalability across heterogeneous meta-atoms. Here, we present a unified large language model (LLM) workflow for multi-family metasurface modeling and inverse-design. Geometries, design parameters, and optical response channels were converted into a shared instruction-following text format and used to fine-tune Gemma-2-9B across 8 metasurface families. Compared with single-family baselines, the joint model simultaneously predicted the optical responses of all metasurface families while reducing the MSE for each family by an average of 56.5\%. The same representation was also used for inverse design. These results show that a shared sequence-based LLM interface can provide a practical route to cross-family metasurface design while reducing the need for task-specific surrogate architectures.
\end{abstract}

\noindent\textbf{Keywords:} metasurfaces; large language models; surrogate modeling; inverse design; nanophotonics

\vspace{\baselineskip}
\noindent *Corresponding Author, E-mail: dhw@psu.edu

\noindent +Second Corresponding Author, E-mail: caozhlong@mail.sysu.edu.cn

\section{1\texorpdfstring{\quad}{ }Introduction}

Metasurfaces have become a central platform for compact optical engineering because subwavelength structuring makes it possible to tailor transmission, reflection, phase, polarization conversion, and resonance behavior in ultrathin devices\cite{r1,r2,r3}. This design freedom has enabled a wide range of applications, including beam shaping\cite{r4}, imaging\cite{r2}, spectral filtering\cite{r5}, chiral response control\cite{r6}, and multifunctional photonic components\cite{r7,r8,r9}. Yet the same geometric freedom that makes metasurfaces attractive also makes them difficult to design efficiently\cite{r10,r11}. Even within a single meta-atom family, one must evaluate complicated mappings from geometric parameters to wavelength-dependent optical responses over a large design space. Standard workflows still rely heavily on repeated numerical simulation\cite{r12,r13}. As metasurface studies move beyond canonical unit-cell families toward broader collections of shapes, materials, and operating conditions, fast models that remain reliable across families have become a practical bottleneck for both scientific exploration and design workflows\cite{r14,r15,r16}.

Data-driven surrogates have emerged as a powerful response to this difficulty\cite{r17,r18,r19,ZhangInverse}. Once trained on geometry-response pairs, neural networks can replace repeated full-wave simulations with near-instant inference and thereby accelerate screening, optimization, and inverse-design loops\cite{r20,r21,r22}. However, most existing surrogates are still organized around a single geometry family or a narrowly defined benchmark\cite{r23,r24}. In practice, a new metasurface class often triggers a new cycle of dataset preparation, network selection, architecture and hyperparameter tuning, loss balancing, and retraining\cite{r16,r17,r25}. Such a workflow is manageable for isolated demonstrations, but it scales poorly when a research program contains several structurally distinct metasurface families that should ideally be studied under one unified modeling framework. For example, a model trained for cylindrical pillars is usually not directly usable for other geometries, such as H-shaped structures\cite{r26}. More importantly, the one-family workflow limits a broader question: whether common electromagnetic patterns learned from one class of metasurfaces can improve prediction on another, rather than forcing each geometry family to remain a separate machine-learning problem.

Recent studies have begun to address this scalability problem through multi-task learning, transfer learning, or generative deep neural networks (DNNs). An \textit{et al.} trained DNNs on full-wave simulation data to achieve scalable objective-driven design across cylindrical, H-shaped, and nearly free-form meta-atoms, but they did not establish one shared model for multiple geometries or metasurface families\cite{r26}. Zhu \textit{et al.} used an ImageNet-pretrained Inception-V3 transfer-learning framework to adapt learning across dielectric-substrate settings of a microwave reflective metasurface, but not across distinct metasurface types\cite{r27}. Yeung \textit{et al.} used a conditional deep convolutional generative adversarial network (cDCGAN) with multiparameter image encoding for global inverse-design across two metasurface families\cite{r28}. However, expanding such image-encoded models to more families may require redesigned representations, more training data, and additional architecture tuning. Zhang \textit{et al.} used heterogeneous transfer learning for data-efficient forward- and inverse-design across several microwave metasurface variants, but they remained within one broad microwave metallic-resonator-on-dielectric-substrate family\cite{r29}. Very recently, Xiao \textit{et al.} reported a powerful machine learning platform for thermal meta-emitters using a human-defined geometric/material descriptor, which enables autoencoder-assisted forward prediction and conditional generative modeling to design meta-emitters within a pre-defined library of 32 types of 3D primitives and 30 candidate materials within one framework\cite{r30}. That framework is versatile, but it remains constrained by predefined primitive and material libraries. OptoGPT formulates multilayer thin-film inverse design as conditional sequence generation using serialized material-thickness tokens\cite{r31}. Together, these studies show important progress toward versatility, but they also expose a central workflow limitation of conventional task-specific DNN surrogates: fixed-dimensional input and output variables, as well as model choices do not naturally accommodate families with different parameter counts, response definitions, spectral ranges, and electromagnetic mechanisms within one shared training interface.

Large Language Models (LLMs) offer a complementary representation because they can encode heterogeneous scientific inputs and outputs through a common sequence-based interface while reducing the amount of task-specific human labor and machine-learning expertise required\cite{r14,r32,r33,r42}. This feature is especially appealing for heterogeneous metasurface families, where geometry classes often differ in parameter definitions, record formats, wavelength ranges, and response conventions. Yet the design tasks still share an underlying requirement to map structural descriptions onto optical behavior. Meanwhile, recent agentic frameworks represent a distinct direction, using LLMs primarily to plan, generate code, and orchestrate learned or physics-based solvers for autonomous photonic design\cite{r34,r35,r36}. Instead of designing a bespoke model architecture for each task, one could represent the relevant properties as structured text and use them to train a single LLM \cite{r14,r32,r33}. An LLM surrogate therefore raises the possibility that one model can absorb cross-family regularities while retaining enough task specificity to produce geometry-aware predictions for multiple metasurface families at once or even perform inverse-design. If successful, such a framework would reduce the need for repeated model engineering, architecture redesign, and human labor when new families are added. This does not remove the need for solid data, correct preprocessing, or physical interpretation, but it can make multi-family surrogate construction more accessible to researchers without specialized machine learning expertise. The unresolved question is whether a single LLM can act as an accurate geometry-aware surrogate and inverse-design engine across heterogeneous metasurface families.

Here, we present a unified LLM workflow for heterogeneous metasurface modeling and inverse-design. Instead of treating each unit-cell family as a separate DNN problem, we express geometry identity, wavelength range, design parameters, and spectral response channels within one shared text representation. This allows one single LLM to learn across distinct metasurface families. Beyond the training methods and trained models, this work also contributes the eight-family instruction-following dataset itself as a reusable training asset. Because the dataset shares the same format, it can be directly merged with future LLM-surrogate corpora. As shown below, several families benefit already upon their first inclusion in joint training and their performance can be further improved as additional families are added. Whether an entirely new future family shows the same benefit remains to be tested. Using this concept, we first trained an LLM for forward predictions and compared isolated single-family training with progressively broader multi-family training. We then used the same concept for inverse-design, where the model predicts geometry parameters from target spectra and a specified geometry family. Together, these results establish that an LLM can provide a practical, simple, and scalable route toward cross-family metasurface learning without the need for task-specific model-architecture and training-pipeline redesign for every new task. We note that in this work, ``toward a universal solver'' refers to an extensible modeling framework in which heterogeneous metasurface families can share one representation, model architecture, and training pipeline, rather than to predict a family absent from fine-tuning.

\section{2\texorpdfstring{\quad}{ }Methods}

\subsection{2.1 Dataset construction}

The study included 8 geometry families in both forward and inverse training. Five families were taken from published datasets, namely cylinder\cite{r26}, H\cite{r26}, AOPS\cite{r37}, micropyramids\cite{r38}, and ellipse\cite{r32}. Three additional in-house families, 4rods, 3rods, and 2rods, were generated from different numbers of silver nanorods embedded in glass using a quasinormal-mode (QNM) simulation workflow (Supplementary Note 1). These families span different structural parameterizations, working frequencies, and optical responses (Figure 1a). The training records were deliberately kept heterogeneous in both design-parameter structure and response-channel schema. Across families, the number of geometric parameters, the number of values in T1 and T2, and the physical meanings assigned to T1 and T2 can all differ. We preserved these differences so that the model had to use the geometry label and full input context to infer the correct family-specific output structure, instead of relying on a fixed-length response vector or manual padding. To train one language model across these systems, each example was converted into a shared instruction-following JSON format, as shown in Supplementary Note 2.

In each json object, the ``geometry'' field stores the assigned family label, and the ``wavelength'' field stores the wavelength range associated with that sample. Most families use one wavelength range within the family, whereas micropyramids use sample-specific ranges that were retained during conversion. The ``parameter'' field stores integer-scaled versions of the original geometric parameters. We did not force all parameters into a single physical unit system, either within a family or across families; instead, numerical magnitudes were only rescaled to reduce token length while retaining the original parameter ordering and family context. The ``T1'' and ``T2'' entries in the ``output'' field are shared channel names, not globally fixed physical observables. Their physical meaning and number of values are defined by the geometry family and the source dataset. For example, T1 and T2 for the ``cylinder'' family represent the real and imaginary parts of the transmission coefficient, each with 61 values, whereas T1 and T2 for the 2rods family represent transmittance and reflectance, each with 41 values. The same magnitude-scaling strategy was applied across these heterogeneous response channels, but the family label and wavelength metadata were retained so the model could infer both the correct channel meaning and the correct output length. Before integer conversion, numerical response values were rounded to three decimal places.

Training and test sets were built by merging and shuffling the selected family datasets. We inherited the train-test split when a source dataset provided one; otherwise, we used an 80:20 random split. The final 8-family dataset contains 392,736 original training samples and 86,039 test samples. Because the training samples for different families were highly imbalanced, the balanced forward training set was constructed by exact duplication of minority-family training samples to bring more equal influence during training, producing 811,422 effective training records for the 8-family run. The test set was never duplicated. We therefore treat the enlarged count as an effective training-record count rather than as additional simulation data. The effect of this exact-duplication balancing was evaluated directly by comparing balanced and unbalanced 8-family forward trainings in Supplementary Figure 3. The same object-level schema also makes the dataset modular: future geometry families can be added by converting their samples to the same format, merging them to the dataset, and rebalancing family counts when needed.

\subsection{2.2 Gemma fine-tuning and forward evaluation}

These datasets were later used to LoRA fine-tune Gemma-2-9B-it-bnb-4bit, a quantized instruction-tuned Gemma-2-9B model. This model was selected based on our previous benchmark of open-weight LLMs for a similar metasurface prediction task\cite{r14}, where this model provided the best overall trade-off between prediction accuracy and computational resource requirements. All fine-tuning processes were performed using Unsloth. In the forward prediction task, the user provides the geometry token, wavelength range, and parameter list, and the model returns the two response arrays in the same JSON format and length expected for that geometry (Figure 1b). A prediction was counted as a parsing failure if the model output could not be parsed into the required fields, if T1 or T2 had the wrong length, or if the predicted numerical arrays remained unusable after the same cleanup rules applied during evaluation. This situation was reported in similar works and can often be solved by letting the model regenerate the output\cite{r14,r33}.

All fine-tuning runs were performed on 4 NVIDIA L40S GPUs in parallel. The per-device batch size was 300, giving an effective batch size of 1200. The LoRA rank $r$ and the scaling factor $\alpha$ were both set to 64, with zero LoRA dropout. We used AdamW with an initial learning rate of $4.0 \times 10^{-4}$, weight decay of 0.01, output-only cross-entropy loss, and linear learning rate decay for a total epoch of 20 but we early stop the training at epoch 9. The forward maximum sequence length was 578 tokens, corresponding to approximately 7.1\% of the 8,192-token native context window of Gemma-2-9B. We trained two groups of forward models. The first group consisted of 8 single-family baselines, one LLM model per geometry, which represents the classical single-task surrogate (Figure 1c, Section 2, Stage 1). The second group consisted of cumulative multi-family models containing two through eight geometry families. At each cumulative step, the included datasets were re-merged, re-balanced, and used to fine-tune one Gemma model (Figure 1c, Section 2, Stage 2-8). This setup resulted in approximately 44 GB of VRAM usage per GPU, and the fine-tuning process took about 137 hours for the dataset with 8 metasurface families. This training time is approximately linearly related to the number of training samples, and the time to generate one prediction is about 0.2 s.

To better illustrate the advantage of multi-family models, we let the single-family models finish fine-tuning for 20 epochs, and the best checkpoint was selected by a score that prioritized MSE while penalizing parsing failure:

\begin{equation}
S_c=\frac{m_c}{M}+0.25\left(\frac{p_c}{P_s}\right)^2+10.00\left[\max\left(0,\frac{p_c}{P_h}-1\right)\right]^2
\end{equation}

where $m_c$ is the checkpoint MSE, $p_c$ is the parsing-failure percentage, $M = 5.0 \times 10^{-3}$, $P_s = 0.5$, and $P_h = 0.8$. This score makes MSE the primary selection criterion, adds a mild preference for lower parsing failure below the soft limit, and applies a stronger penalty when parsing failure exceeds the hard limit. Because the single-family baselines were selected from their best checkpoint histories while the multi-family runs had a common epoch, the comparison is favorable to the single-family baselines and conservative with respect to the multi-family models. Later, we show that even under this ``unfair'' comparison, the multi-family models can also beat the single-family models.

\begin{center}
\includegraphics[width=\textwidth]{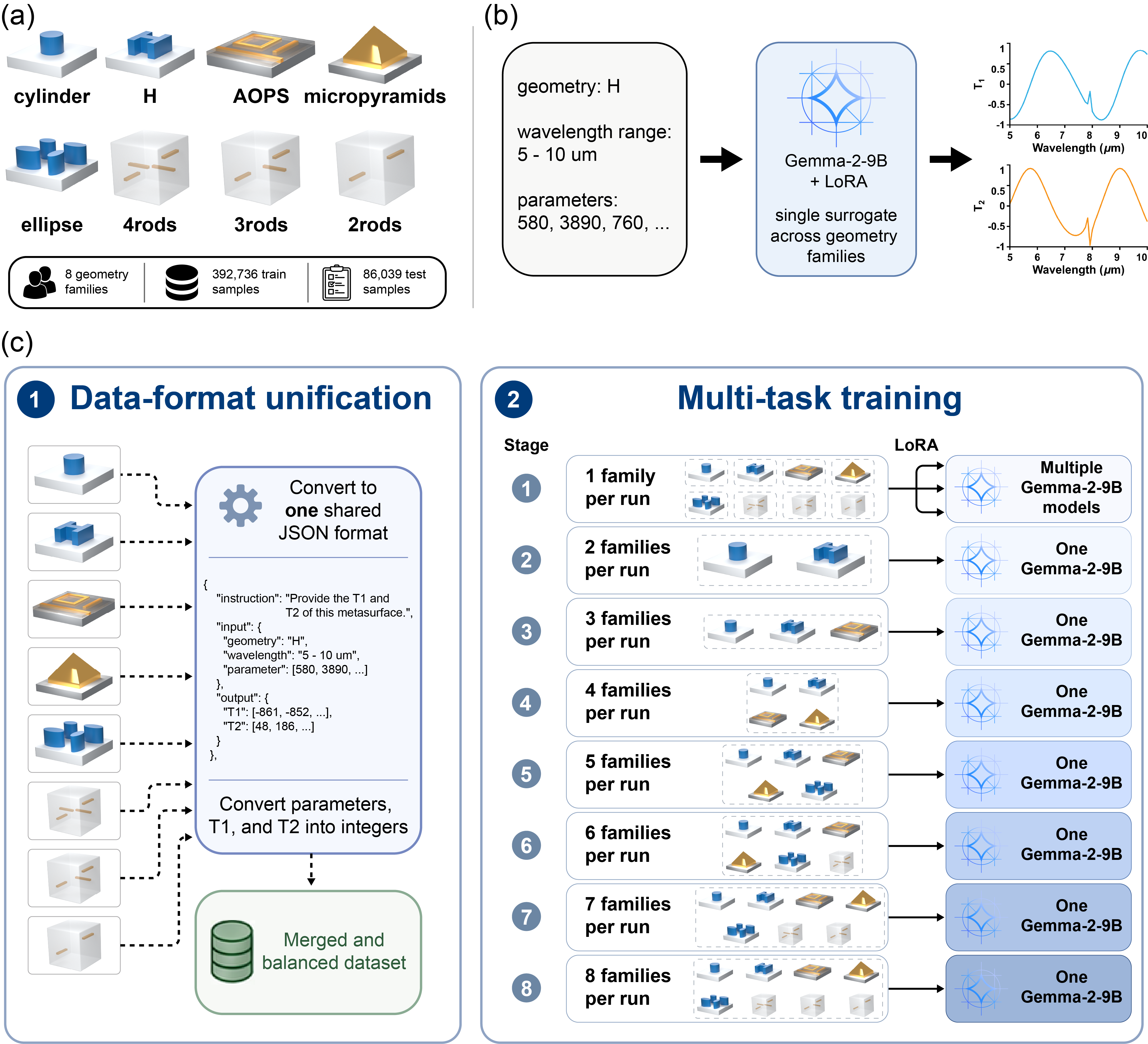}
\end{center}
\noindent\textbf{Figure 1. Unified language-model workflow for multi-family metasurface prediction.} (a) The 8 geometry families included in the benchmark. (b) Forward prediction task: a JSON description of geometry, wavelength range, and parameters is provided to a LoRA fine-tuned Gemma-2-9B model, which returns T1 and T2 response arrays. (c) Data-format unification and cumulative multi-task training workflow.

\subsection{2.3 Inverse-design and round-trip evaluation}

The inverse-design task used the same representation with the input-output direction reversed. The input contains the target spectra, geometry token, and wavelength range, and the output contains one parameter list, as shown in Supplementary Note 2.

The inverse model was trained separately from the forward benchmark on the merged 8-family inverse dataset. It used the same fine-tuning hyperparameters and strategies as the forward model, with a maximum sequence length of 583 tokens and was fine-tuned for 15 epochs. As discussed previously\cite{r14}, inverse metasurface design is generally nonunique and an LLM mitigates the need to utilize some unusual training techniques such as tandem network, as the LLM internally supports multiple outputs for the same input. For the same reason, direct parameter error is not the main quality metric. Instead, we evaluated inverse predictions by round-trip spectral agreement (Figure 4a). First, the inverse model predicted a parameter list from target spectra and geometry metadata. Second, the predicted parameters were passed to the fine-tuned joint 8-family forward model for non-rod geometries. Third, the resulting spectra were compared with the target T1 and T2 arrays by MSE. Such round-trip evaluation is widely used in tandem or bidirectional neural-network inverse design of metasurfaces\cite{r39,r40,r41}. For the rod families, the epoch-15 test-set MSE was computed from QNM responses for the entire rod test set, whereas the rod examples displayed in Figure 4 were additionally simulated with COMSOL for one representative design from each rod family. Parsing failure in the inverse workflow includes inverse parameter-output failures and fixed-forward output failures.

\section{3\texorpdfstring{\quad}{ }Results and discussion}

\subsection{3.1 Forward prediction with the joint 8-family model}

The joint 8-family forward model achieved a global MSE of $1.1 \times 10^{-3}$ and a global parsing failure rate of 0.292\% for predicting the merged 8-family test set. Figure 2a shows that the average MSE decreased rapidly during the first few epochs and then improved more gradually through the training process. The MSE and parsing failure rate for each family are listed in the caption of Figure 2.

Representative spectra in Figure 2c-j compare ground-truth responses, predictions from the joint 8-family model, and predictions from the corresponding single-family baseline model. The orange dashed curves from the joint 8-family model closely track the blue solid ground truth curves across each family-specific wavelength band, including plateaus, broad resonances, and sharp spectral features. The relative benefit varies by family and by sample, but these examples demonstrate that the multi-family model often tracks spectral features more closely than the single-family baseline. To check that this behavior was not limited to only the selected spectra in Figure 2, Supplementary Figure 2 compares the full MSE distributions of the joint 8-family model and the corresponding single-family baselines for each geometry family. In other words, a lightly fine-tuned, off-the-shelf LLM delivers turnkey, high-fidelity forward modeling and, more importantly, shows versatility across multiple metasurface families without any bespoke network design or hyperparameter sweeps, demonstrating the potential of this extensible multi-family framework as a step toward a universal meta-optics solver.

\begin{center}
\includegraphics[width=\textwidth]{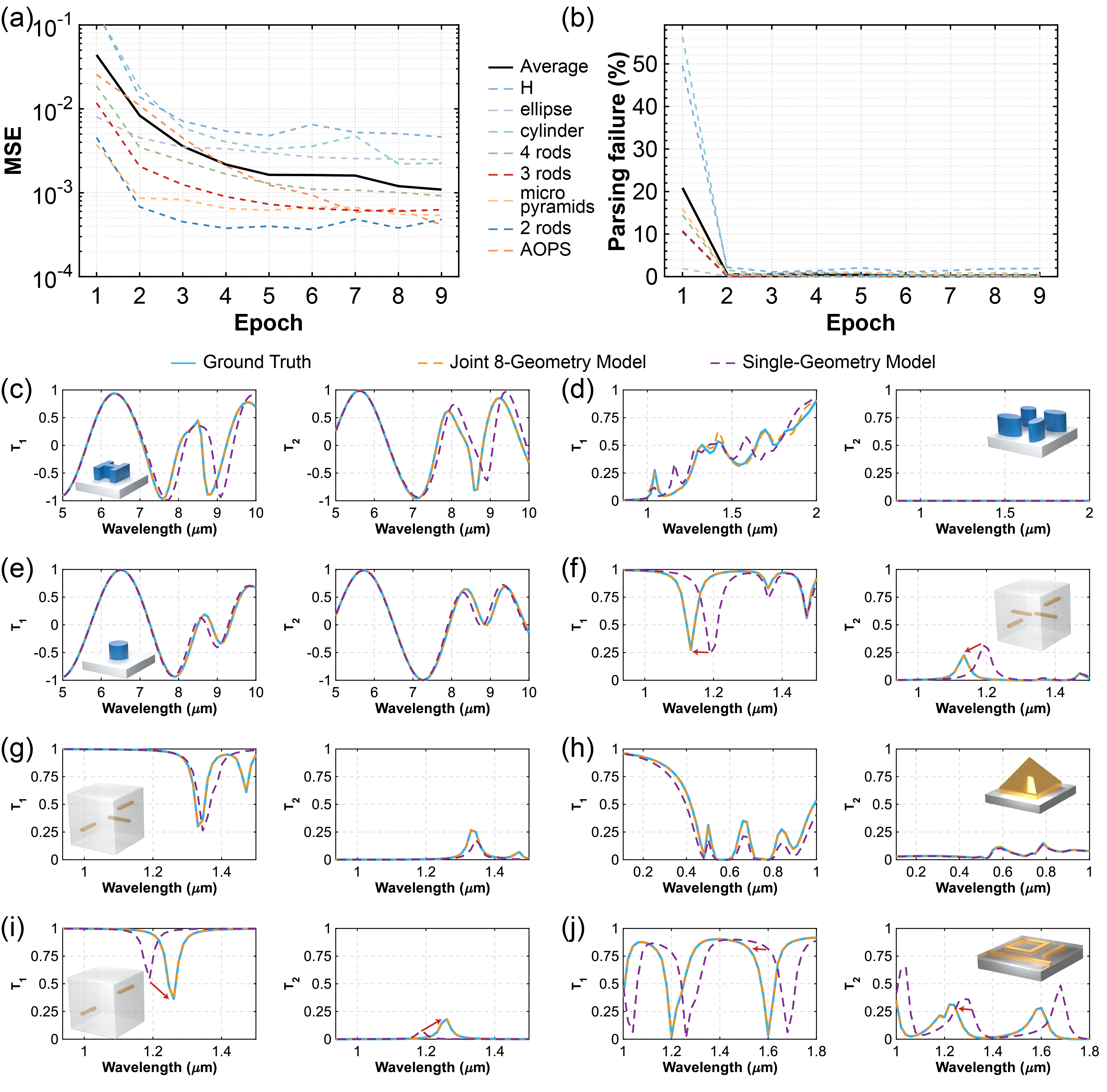}
\end{center}
\noindent\textbf{Figure 2. Forward prediction using the joint 8-geometry model.} (a) Geometry-resolved and average MSE versus epoch for the 8-family forward run. The final MSE values for each family are: H: $4.6 \times 10^{-3}$; ellipse: $2.5 \times 10^{-3}$; cylinder: $2.2 \times 10^{-3}$; 4 rods: $9.2 \times 10^{-4}$; 3 rods: $6.3 \times 10^{-4}$; micropyramids: $5.4 \times 10^{-4}$; 2 rods: $4.8 \times 10^{-4}$; AOPS: $4.1 \times 10^{-4}$. On average, the final MSE for all test samples is $1.1 \times 10^{-3}$. (b) Geometry-resolved and average parsing-failure rate versus epoch. The final parsing failure rates for each family are: H: 1.90\%; ellipse: 0.10\%; cylinder: 0.34\%; 4 rods: 0.06\%; 3 rods: 0.09\%; micropyramids: 0.63\%; 2 rods: 0.26\%; AOPS: 0.13\%. On average, the final parsing failure rate for all test samples is 0.29\%. (c-j) Representative forward spectra comparing ground truth, the joint 8-geometry model, and the corresponding single-geometry baseline model for different metasurface families.

\subsection{3.2 Multi-family training improves per-family accuracy}

The forward results demonstrate that a single model can simultaneously predict the optical responses of multiple metasurface families while improving the prediction accuracy for individual families by leveraging knowledge shared across families. Figure 3 further answers whether adding heterogeneous families can improve prediction for all individual families included within the same LLM training framework. Figures 3a,b show the trend of MSE and parsing-failure rate as the number of families in the cumulative training run increases. Blank (white) regions indicate that a geometry family was not yet present in that cumulative run. The first populated cell for each later-added family represents its performance when it is first incorporated into a multi-family training mixture. For example, the two-family run contains only H and cylinder, while later runs add more families until the 8-family run includes all benchmark geometries. Figures 3a and b illustrate one cumulative-training path using a fixed family-addition sequence. The intermediate results are not intended to establish the relationship between invariance and the adding order of the family. For most of these families, this initial joint-training result has been already improved upon the corresponding single-family baseline. Thus, Figure 3 shows two related effects within the present benchmark: a family can benefit upon its inclusion in joint training, and families already included can benefit further as more families are introduced, although improvements are not strictly monotonic for every family.

The bar charts in Figures 3c,d compare each optimal single-family baseline metric with the joint 8-family model for the same geometry. The 8-family models reduce MSE for all metasurface families in this comparison. The average relative MSE reduction is 56.5\%. The largest reductions occur for 2rod and H structures, with MSE reductions of about 90\%. In the worst case, for the micropyramids family, the MSE still improves by 5\%. Notably, for the ellipse family, where the original study also uses an LLM (ChatGPT) for optical response prediction\cite{r32}, our training strategy achieves an even lower prediction MSE than their engineered approach (Supplementary Table 1). These results indicate that the benefit depends on the family and its baseline strength. Parsing-failure changes were much smaller because most models already returned valid structured outputs. The average reduction is 0.6 percentage points. The H family shows the largest parsing-failure improvement, decreasing by 3.6 percentage points relative to its single-family baseline, whereas the ellipse shows a slight 0.1 percentage-point increase.

\begin{center}
\includegraphics[width=\textwidth]{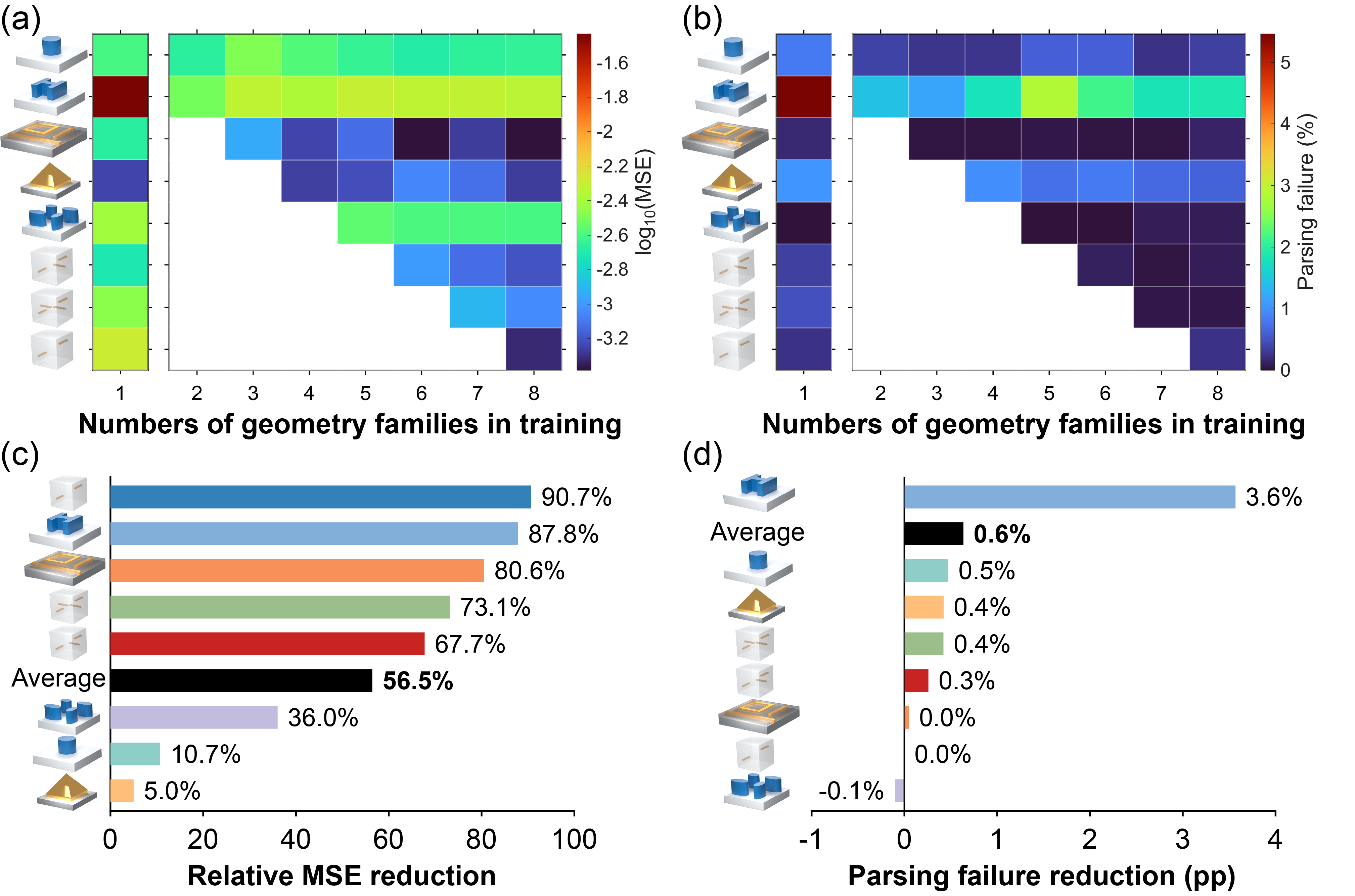}
\end{center}
\noindent\textbf{Figure 3. Effect of cumulative multi-family training.} (a) Heat map of MSE values across geometry families and cumulative training runs. (b) Heat map of selected parsing-failure rates. (c) Per-family relative MSE reduction of the joint 8-family model relative to the selected single-family baseline. (d) Parsing failure rate reduction in percentage points (pp) relative to the selected single-family baseline. Positive values indicate fewer parsing failures for the joint model.

These results support two distinct claims within the scope of our study in this work. The directly measured claim is that, within the LLM framework, multi-family training can improve numerical accuracy relative to isolated single-family fine-tuning, especially for families whose single-family baselines are weaker or whose datasets benefit from cross-family regularities. The broader workflow claim is that the shared sequence representation reduces the need to build a new descriptor, architecture, output layer, and training pipeline for every metasurface family. We therefore do not interpret cross-paper reported MSE values from prior DNN studies as direct accuracy benchmarks. Instead, the advance demonstrated here is versatility under one modeling interface, together with positive cross-family transfer in the controlled LLM comparison, as well as reduced entry barriers and human labor. Together, the cumulative-family experiments suggest a practical reuse route for future geometries. After converting a new geometry family into the same instruction-following schema, the training samples can be merged with the dataset and used for joint fine-tuning. The comparisons in Figure 3 and Supplementary Figure 3 suggest that, when the new family shares learnable structure with the existing corpus, this joint training would be expected to reduce the geometry-specific MSE relative to an isolated single-family run, but quantitative benefit remains to be tested.

\subsection{3.3 Inverse-design}

The inverse task was trained separately from the forward benchmark and evaluated by round-trip spectral agreement. In this task, the input contains T1, T2, the geometry token, and the wavelength range, while the output contains only a parameter list. As discussed in the Methods section, we evaluated the inverse model using the workflow outlined in Figure 4a. Figure 4b shows that the inverse round-trip MSE improved during the early epochs and then reached a broad plateau. At epoch 15, 85,755 of 86,039 test samples produced valid round trips. Figures 4c-j show representative inverse-designed spectra. The global round-trip MSE for non-rod families is $2.97 \times 10^{-3}$, and the global QNM MSE for 2-4 rod families is $4.05 \times 10^{-4}$. Combining these groups gives a global inverse-design MSE of $2.06 \times 10^{-3}$. For the non-rod families, the reported MSE should be interpreted as a surrogate-based evaluation metric, since correlated forward- and inverse-model errors cannot be excluded. These inverse results show that the same language-based representation can support not only forward prediction but also inverse-design across heterogeneous metasurface families. They also demonstrate a practical multi-family design loop: one LLM predicts candidate parameters from target spectra, and a fixed evaluator checks whether those parameters reproduce the desired optical response.

\begin{center}
\includegraphics[width=\textwidth]{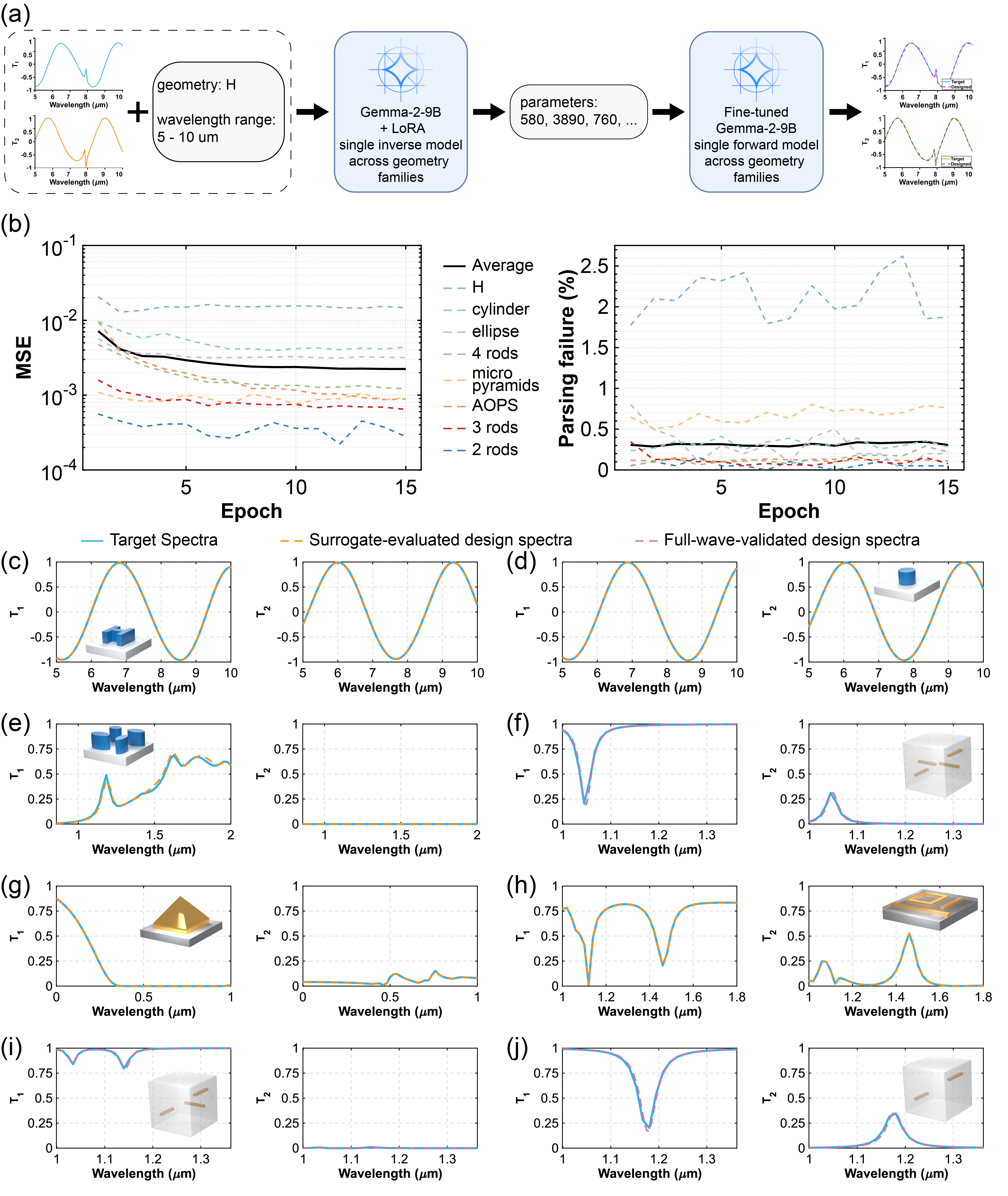}
\end{center}
\noindent\textbf{Figure 4. Inverse-design evaluated by round-trip spectral agreement.} (a) Inverse evaluation workflow. Target spectra and geometry metadata are passed to a single fine-tuned inverse Gemma-2-9B model, which predicts geometry parameters. The predicted parameters are then evaluated by a fine-tuned joint 8-family forward model or by QNM, and the designed spectra are compared with the target spectra. (b) Inverse round-trip MSE and combined parsing-failure rate versus epoch. The round-trip or QNM MSE values for each family are: H: $1.5 \times 10^{-2}$; cylinder: $4.4 \times 10^{-3}$; ellipse: $3.2 \times 10^{-3}$; 4 rods: $5.6 \times 10^{-4}$; micropyramids: $9.2 \times 10^{-4}$; AOPS: $8.9 \times 10^{-4}$; 3 rods: $2.7 \times 10^{-4}$; 2 rods: $2.3 \times 10^{-4}$. On average, the MSE for all test samples is $2.06 \times 10^{-3}$. The parsing failure rates for each family are: H: 1.87\%; cylinder: 0.28\%; ellipse: 0.20\%; 4 rods: 0.22\%; micropyramids: 0.76\%; AOPS: 0.11\%; 3 rods: 0.08\%; 2 rods: 0.05\%. On average, the parsing failure rate for all test samples is 0.33\%. (c-j) Representative inverse-designed spectra for the 8 geometry families.

\section{4\texorpdfstring{\quad}{ }Conclusions}

This work demonstrates a unified LLM workflow for heterogeneous metasurface modeling and inverse-design. By representing geometry identity, wavelength range, design parameters, and optical responses in a shared instruction-following format, a single LoRA fine-tuned Gemma-2-9B model can learn across 8 structurally distinct metasurface families without family-specific input layers, output heads, or descriptor redesign. Although this study considers eight metasurface families, the proposed framework is not inherently limited to this number. Given sufficient high-quality training data, it should be scalable to a considerably broader range of metasurface geometries. Finer spectral sampling, longer parameter lists, or additional response channels can in principle be accommodated within the same representation up to the model context limit, although longer sequences increase computational and memory costs and their performance remains to be tested.

Furthermore, the forward benchmark shows that multi-family training can improve accuracy beyond isolated single-family fine-tuning. The joint 8-family model not only shows versatility but also reduces MSE for every family relative to the selected single-family baselines. The average relative MSE reduction was 56.5\%, with the largest gains observed for H structures and 2rod structures. The inverse-design results extend the same sequence-based interface from response prediction to parameter generation.

Together, these results support the use of LLMs as flexible surrogate engines for multi-family meta-optics workflows. The main advantage is not a direct claim of universal accuracy across all metasurfaces, but a reusable text-based modeling interface that can absorb heterogeneous datasets, compare single-family and multi-family training under one framework, and support forward and inverse tasks with minimal architecture redesign. The dataset produced for this study is also a reusable contribution: its instruction-following format allows future geometry-specific datasets to be merged without requiring changes to the model architecture. Future work could test previously unseen families, extrapolation beyond trained parameter ranges, broader material and structural settings, and more diverse response representations, further extending the present framework toward a universal, general-purpose meta-optics solver.

\section{Disclosures}

The authors declare that there are no financial interests, commercial affiliations, or other potential conflicts of interest that could have influenced the objectivity of this research or the writing of this paper.

\section{Code, Data, and Materials Availability}

Code and data used in this work can be found on\\
\url{https://github.com/zhanghsh9/MultipleShapeLLM}.

\section{Acknowledgments}

YC, LW and ZC acknowledge support provided by National Key Research and Development Program of China (2022YFA120660). HZ, LK, and DHW are grateful for the support of this work from The Pennsylvania State University John L. and Genevieve H. McCain Endowed Chair Professorship.

\bibliography{manuscript}

\clearpage
\includepdf[pages=-,pagecommand={}]{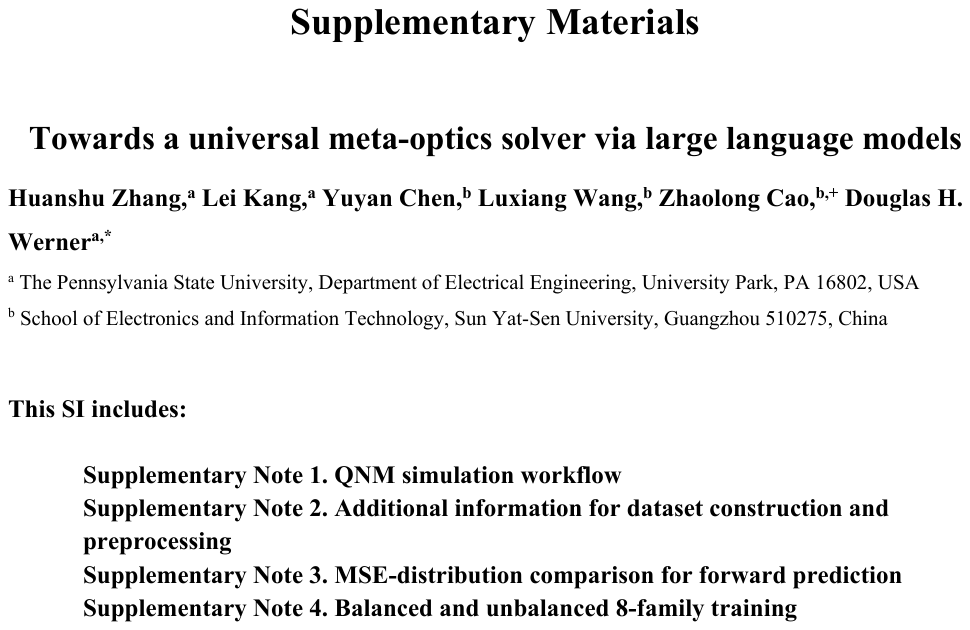}

\end{document}